\documentclass[journal=jacsat,manuscript=article]{achemso}

\usepackage{graphicx}
\usepackage{dcolumn}
\usepackage{bm}

\usepackage[utf8]{inputenc}
\usepackage{mathptmx}
\usepackage{multirow}
\usepackage{braket}

\usepackage[english]{babel}
\usepackage{xcolor}
\colorlet{RED}{red}
\colorlet{BLUE}{blue}
\usepackage{dcolumn}
\usepackage{bm}
\usepackage[version=4]{mhchem} 
\usepackage{acronym}
\usepackage{adjustbox}
\usepackage{float}
\usepackage{tikz}
\usepackage{microtype} 
\usepackage{algorithm}
\usepackage{xcolor}
\usepackage{algpseudocode}
\usepackage{comment}

\newcommand{\onlinecite}[1]{\hspace{-1 ex} \nocite{#1}\citenum{#1}}

\usetikzlibrary{calc,shapes.geometric,decorations.pathmorphing,patterns}

\definecolor{background-color}{gray}{0.98}

\title{
Density Matrix Renormalization Group Approach Based on the Coupled-Cluster Downfolded Hamiltonians
}

\author{Nicholas Bauman}
\email{nicholas.bauman@pnnl.gov}
\affiliation{
  Physical Sciences Division, 
  Pacific Northwest National Laboratory, Richland, Washington, 99354, USA
}

\author{Libor Veis}
\email{libor.veis@jh-inst.cas.cz}
\affiliation{
J. Heyrovsky Institute of Physical Chemistry, v.v.i., Czech Academy of Sciences, Prague, Czechia
}

\author{Karol Kowalski}
\email{karol.kowalski@pnnl.gov}
\affiliation{
  Physical Sciences Division, 
  Pacific Northwest National Laboratory, Richland, Washington, 99354, USA
}

\author{Jiri Brabec}
\email{jik.jiri@gmail.com}
\affiliation{
J. Heyrovsky Institute of Physical Chemistry, v.v.i., Czech Academy of Sciences, Prague, Czechia
}

\begin{document}

\begin{abstract}
The Density Matrix Renormalization Group (DMRG) method has become a prominent tool for simulating strongly correlated electronic systems characterized by dominant static correlation effects. However, capturing the full scope of electronic interactions, especially for complex chemical processes, requires an accurate treatment of static and dynamic correlation effects, which remains a significant challenge in computational chemistry. This study presents a new approach integrating a Hermitian coupled-cluster-based downfolding technique, incorporating dynamic correlation into active-space Hamiltonians, with the DMRG method. By calculating the ground-state energies of these effective Hamiltonians via DMRG, we achieve a more comprehensive description of electronic structure. We demonstrate the accuracy and efficiency of this combined method for advancing simulations of strongly correlated systems using benchmark calculations on molecular systems, including N$_2$, benzene, and tetramethyleneethane (TME).
\end{abstract}

\maketitle


The availability of predictive electronic structure methods capable of dealing with complex correlation effects will play a pivotal role in further advancing chemistry and material science. Among many candidates based on various representations of quantum mechanics, two wave-function-based methodologies
coupled cluster (CC) formalism \cite{coester58_421,coester60_477,cizek66_4256,paldus1972correlation,purvis82_1910,paldus07,crawford2000introduction,bartlett_rmp} and density matrix renormalization group (DMRG) formalisms have assumed unique positions in computational chemistry. 


The single-reference coupled cluster (SR-CC) formalism \cite{coester58_421,coester60_477,cizek66_4256} 
serves as a theoretical framework for systematically capturing various forms of correlation effects. Initially intended for modeling closed-shell systems dominated by dynamic electronic correlation, the CC theory has been extended to address more complex scenarios involving the coexistence of dynamic and significant static correlation effects by including higher-rank excitations and multi-reference extensions. Tackling strong correlation effects continues to be an area of active development, with ongoing efforts focused on translating new concepts into algorithms that leverage emerging computational technologies such as high-performance computing (HPC) and quantum computing (QC). Moreover, there has been a significant push towards developing hybrid computational approaches that combine the strengths of HPC and QC to enhance the accuracy of CC many-body formulations.

The Density Matrix Renormalization Group (DMRG) offers a computational algorithm for optimizing the wave function within the problem-specific active space (AS) by utilizing singular value decomposition (SVD) steps to factorize coefficients in the configuration-interaction-type expansion. One advantage of the DMRG method is its agnosticism toward the complexity of the wave function in the active space, as it does not rely on any specific form of Ansatz or reference function, as seen, for example,  in the SR-CC approach. Consequently, DMRG can effectively describe multiconfigurational systems when the sought-after ground state's structure is complex and unknown before the simulations. However, while DMRG can yield meaningful results regarding the ordering of electronic states or energy differences between states of various multiplicities in many applications, its focus solely on recovering static correlation effects may adversely impact prediction quality, particularly in situations where the structure of the states of interest strongly depends on the geometry of the chemical system. Several formulations have been developed to integrate DMRG with perturbation theory and coupled cluster formalisms to address these issues \cite{neuscamman_2010_irpc, Roemelt2016, Freitag2017, sharma_2014c, Veis2016, Beran2021}.

Recently,  CC downfolding techniques have emerged to address the issue of reducing the dimensionality of quantum problems. In these methodologies, the exponential CC Ansatz plays a crucial role in constructing active-space effective Hamiltonians, ensuring that the lowest eigenvalue of the effective Hamiltonian aligns with the exact or approximate CC energy. CC downfolding allows for the incorporation of dynamical correlation effects into the many-body structure of the effective Hamiltonian. Furthermore, these techniques encompass two types of downfolding associated with applying SR-CC and general unitary CC Ansatzen. The latter, known as Hermitian CC downfolding (HDCC), holds particular significance in quantum computing (QC) applications, although various solvers can be used to diagonalize the HDCC effective Hamiltonian. Moreover, non-Hermitian CC downfolding presents an elegant approach to mathematically integrate different active space problems (see the Equivalence Theorem of  Refs.~\onlinecite{kowalski2021dimensionality,bauman2022coupled,kowalski2023quantum}) thereby giving rise to the concept of quantum flow. 
Due to the factors above, we aim to assess the effectiveness of the integrated HDCC and DMRG framework in capturing dynamical correlation effects.






In what follows, we briefly describe the salient features of the DMRG and CC downfolding methods.  
The DMRG method is a highly accurate numerical technique used for studying strongly correlated quantum systems  \cite{schollwock_2005}. It works by optimizing the representation of a many-body wave function in a truncated Hilbert space, significantly enhancing computational efficiency while maintaining a high degree of precision. In quantum chemistry \cite{White1999, chan_review, wouters_review, Szalay2015, reiher_perspective}, the approximated wave function typically represents the ground state (or low-lying excited states) of a full configuration interaction (FCI) solution in a specified orbital space, such as the complete active space configuration interaction (CASCI) framework. The DMRG algorithm \cite{White1992, White-1993} recasts the wave function into a matrix product state (MPS) representation, providing an efficient and compact form for describing entangled quantum states \cite{Schollwock2011}.

The FCI wave function, in the occupation number basis, is expressed as
\begin{equation}
  | \Psi_{\text{FCI}} \rangle = \sum_{\{\alpha\}} c^{\alpha_1 \alpha_2 \ldots \alpha_n} | \alpha_1 \alpha_2 \cdots \alpha_n \rangle,
\end{equation}
where $\alpha_i$ represents the occupation state of the $i$-th orbital, with $\alpha_i \in { | 0 \rangle, | \downarrow \rangle, | \uparrow \rangle, | \downarrow \uparrow \rangle }$. By successively applying SVD to the FCI tensor $c^{\alpha_1 \alpha_2 \ldots \alpha_n}$, the wave function can be factorized into an MPS form \cite{Schollwock2011}
\begin{equation}
  \label{mps_factorization}
  c^{\alpha_1 \ldots \alpha_n} = \sum_{i_1 \ldots i_{n-1}} A[1]_{i_1}^{\alpha_1} A[2]_{i_1 i_2}^{\alpha_2} A[3]_{i_2 i_3}^{\alpha_3} \cdots A[n]_{i_{n-1}}^{\alpha_n},
\end{equation}
where $A[j]^{\alpha_j}$ are the MPS matrices (with $A[1]^{\alpha_1}$ and $A[n]^{\alpha_n}$ being vectors rather than matrices) corresponding to each orbital. For clarity, we omit the $[j]$ index notation and simply refer to the MPS matrices as $A^{\alpha_j}$. The new indices, $i_j$, introduced by SVD, are called virtual indices, and they are contracted across different MPS matrices. If the MPS factorization was exact, the dimensions of these matrices would grow exponentially with a system size, similar to the growth of the original FCI tensor. In the DMRG algorithm, however, the dimensions of the virtual indices are deliberately truncated and bounded, resulting in reduced computational complexity. These dimensions are known as bond dimensions, typically denoted by $M$. The choice of $M$ controls the accuracy of the approximation, with larger bond dimensions capturing more entanglement at the cost of greater computational effort.

A practical two-site DMRG algorithm provides the wave function in the two-site MPS form
\begin{equation}
  \label{eq:MPS_2site}
  | \Psi_\text{MPS} \rangle = \sum_{\{\alpha\}} \mathbf{A}^{\alpha_1} \mathbf{A}^{\alpha_2} \cdots \bm{\psi}^{\alpha_j \alpha_{j+1}} \cdots \mathbf{A}^{\alpha_n}| \alpha_1 \alpha_2 \cdots \alpha_n \rangle,
\end{equation}
where \( \bm{\psi}^{\alpha_j \alpha_{j+1}}_{i_{j-1}i_{j+1}}\) stands for the four-index tensor representing the eigenstate of the electronic Hamiltonian
\begin{equation}
H_\mathrm{el.} = \sum_{\sigma} \sum_{pq} h_{pq} a_{p_{\sigma}}^{\dagger} a_{q_{\sigma}} +
    \frac{1}{2} \sum_{\sigma \sigma^{\prime}}\sum_{pqrs} \langle pq | rs \rangle a_{p_{\sigma}}^{\dagger} a_{q_{\sigma^{\prime}}}^{\dagger} a_{s_{\sigma^{\prime}}} a_{r_{\sigma}},
  \label{ham_sec_quant}
\end{equation}
expanded in the tensor product space of four distinct vector spaces defined along an ordered orbital chain: the left block (orbitals $1 \ldots j-1$), the two explicit sites ($j, j+1$), and the right block ($j+2 \ldots n$). In Eq. \ref{ham_sec_quant}, $h_{ij}$ and $\langle ij | kl \rangle$ denote standard one and two-electron integrals in the molecular orbital (MO) basis, and $\sigma$ and $\sigma^{\prime}$ denote spin, $\sigma, \sigma^{\prime} \in \{ \uparrow, \downarrow \}$.
The optimization of MPS matrices is performed sequentially in a process called sweeping. 

In the original DMRG formulation \cite{White1992, White-1993, schollwock_2005}, the explicit (determinant) representations of the complicated many-particle bases 
are not stored, but the matrix representations of second-quantized operators needed for the action of the Hamiltonian on a (trial) wave function 
are formed and stored. 
The transition between iterations in a DMRG sweep is achieved through a renormalization procedure. During this step, the basis of one block, in the direction of the sweep, is expanded by adding a new site, while the operators required to apply the Hamiltonian to the wave function are transformed into this new basis. Meanwhile, the complementary block is reduced by removing one site. This process effectively shifts the position within the MPS by one matrix, allowing the diagonalization of the effective Hamiltonian 
to be repeated.

The key part of the DMRG algorithm - the truncation - is done in such a way, that the density matrix of the enlarged block changes as little as possible. This is achieved by diagonalizing the density matrix and retaining only the largest eigenvalues. 

An alternative to forming renormalized operators is to factorize the Hamiltonian (Eq.~\ref{ham_sec_quant}) into a Matrix Product Operator (MPO) form \cite{Schollwock2011, keller_2015, keller_2016}. In this approach, the effective Hamiltonian corresponds to a partially contracted tensor network. Both methods are equivalent in terms of computational efficiency and there is a clear transition between them \cite{chan_mpo}.

Two important correlation measures that play a significant role in tuning the performance of DMRG are the single-orbital entanglement entropy (\( s_i \)) and the mutual information (\( I_{ij} \)) \cite{legeza_2003b,rissler_2006,barcza_2011}. The single-orbital entanglement entropy quantifies the importance of an orbital in the wave function expansion and is computed as
\begin{equation}
  s_i = -\text{Tr} \rho_i \ln \rho_i,
\end{equation}
where \( \rho_i \) represents the reduced density matrix of orbital \( i \) \cite{barcza_2015}. When considering a pair of orbitals (\( i \), \( j \)), the two-orbital entanglement entropy, \( s_{ij} \), can be similarly derived. The mutual information is defined as
\begin{equation}
  I_{ij} = s_{ij} - s_{i} - s_{j},
\end{equation}
which quantifies the correlation between orbitals \( i \) and \( j \) within the broader system \cite{rissler_2006,barcza_2011}.


The compression of the dimensionality of the quantum problem in the form of relevant active space has been and continues to be an active area of development.
Recent developments (generally referred to as the  CC downfolding approaches) indicate that the effective Hamiltonian theory is also an inherent feature of standard single reference CC (SR-CC ) formulations, allowing for an alternative way of calculating SR-CC ground-state energies (see Refs.~\onlinecite{kowalski2021dimensionality}, \onlinecite{kowalski2018properties}, and \onlinecite{kowalski2023sub}). 
The Hermitian extension of the CC downfolding\cite{bauman2019downfolding,bauman2022coupled},  which utilizes the so-called double unitary coupled cluster (DUCC) Ansatz, has been introduced in the context of the quantum computing applications.
In analogy to the non-Hermitian SR-CC case, the DUCC Ansatz leads to the many-body form of the active-space Hermitian effective Hamiltonian, $H^{\rm eff}$
\begin{equation}
    H^{\rm eff} = (P+Q_{\rm int})
    e^{-\sigma_{\rm ext}} H
    e^{\sigma_{\rm ext}}
    (P+Q_{\rm int})
    \label{ducc1}
\end{equation}
where $\sigma_{\rm ext}$ is the so-called external anti-Hermitian cluster operator, $P$ is the projection operator onto the reference function $|\Phi\rangle$ (usually chosen as a Hartree--Fock (HF) Slater determinant), and $Q_{\rm int}$ is the projection operator onto excited configurations (with respect to the $|\Phi\rangle$ determinant) belonging to complete active space (CAS) of interest.  

In practical applications, the construction of a second quantized representation of $H^{\rm eff}$ is associated with several approximations. First, the $\sigma_{\rm ext}$ operator is approximated in the unitary CC (UCC) form
\begin{equation}
    \sigma_{\rm ext} \simeq T_{\rm ext} - T_{\rm ext}^{\dagger} \;,
    \label{ducc3}
\end{equation}
where $T_{\rm ext}$ is the external part of the standard CC cluster operator defined by cluster amplitudes that carry at least one inactive spin-orbital index (the partitioning of the SR-CC cluster operator into the internal and external parts originates in the active-space SR-CC theory, see Ref.~\onlinecite{pnl93}). For practical reasons, in our approach, the  $T_{\rm ext}$ is approximated by the external part of the CCSD cluster operator. Second, the expansion (\ref{ducc1}) is non-terminating; therefore, we use the finite-rank commutator expansion stemming from the Baker–Campbell–Hausdorff formula. 
We investigate two approximations of the commutator expansion - a truncation that is consistent through second-order perturbation theory, which we call DUCC(2), and a truncation that is consistent through third-order perturbation theory, which we call DUCC(3)
(see approximations A4 and A7 in Ref.~\onlinecite{doublec2022}, respectfully).
Additionally, the rank of many-body effects included in $H^{\rm eff}$ are limited to the one- and two-body interactions:
\begin{equation}
H^{\rm eff} \simeq
\Gamma_0 +
\sum_{pq} g^p_q a_p^{\dagger} a_q + \frac{1}{4} \sum_{p,q,r,s} k^{pq}_{rs} a_p^{\dagger} a_q^{\dagger} a_s a_r \;,
\label{ducc5}
\end{equation}
where $\Gamma_0$ is a scalar, $g^p_q$ and 
$k^{pq}_{rs}$ tensors define one- and two-body effective interactions, and indices $p$, $q$, $r$, $s$ designate active spin-orbitals.


The HDCC-DMRG workflow utilizes two computational drivers briefly described below: \\
{\bf HDCC implementation:}
The parallel HDCC code calculates the  $\Gamma_0$ scalar and  $g^p_q$/ $k^{pq}_{rs}$ tensors. In the process of implementing this approach, we used the SymGen infrastructure (\url{https://github.com/npbauman/SymGen}), which uses symbolic algebra tools to automatically derive tensor expressions corresponding to the second quantized form of operators encountered in various CC methodologies. The output of SymGen is translated into the parallel tensor contraction library TAMM 
(Tensor Algebra for Many-body Methods)\cite{mutlu2023tamm}
format, which
streamlines the translation of tensor contractions to efficient
parallel code  and eliminates the need for the
hand-derivation of hundreds or thousands of diagrams.
The  HDCC is available in the ExaChem software repository
located at \url{https://github.com/ExaChem/exachem}. Due to the TAMM flexibility in taking advantage of a variety of GPU architectures, the HDCC code (and ExaChem in general) can effectively utilize Leadership Computing Facilities and HPC cloud computing services.\cite{...} \\
{\bf DMRG implementation:} The parallel DMRG implementation (MOLMPS program) was presented elsewhere \cite{molmps}. It currently supports distributed matrix manipulations on both CPUs and GPUs, making it well-suited for use in Leadership Computing Facilities and HPC cloud computing services. The interface between the TAMM library and the MOLMPS program manages communication of \mbox{$g^p_q$/ $k^{pq}_{rs}$} tensors, although these tensors generally exhibit lower symmetry compared to non-relativistic MO integrals. Consequently, the eight-fold permutational symmetry cannot be fully exploited in the internal storage of MOLMPS.


The purpose of the performed calculations is to compare DMRG with DMRG-DUCC methods and analyze the changes in AS correlation patterns. We chose three systems: nitrogen molecule (\(\text{N}_2\)), tetramethyleneethane (TME), and benzene. 
For \(\text{N}_2\), we employed the cc-pVDZ basis set, so we have a comparison with FCI energies obtained by DMRG in full space (28 orbitals). For benzene and TME, we used the cc-pVTZ basis set and focused on the lowest singlet state. For TME, we computed energies for torsion angles 0, 30, 45, 60 and 90 degrees, we used AS consisting of all occupied orbitals (22) and 12 virtual orbitals - CAS(44,34). The SCF, CCSD, DUCC(2) and DUCC(3) calculations were performed by TAMM software, while DMRG, DMRG-DUCC and the entanglement analysis using the MOLMPS program.\cite{molmps}\\ 

\begin{table}[htbp]
\centering
\small 
\setlength{\tabcolsep}{4pt} 
\renewcommand{\arraystretch}{1.1} 
\begin{tabular}{|l|ccc|ccc|}
\hline
 & \multicolumn{3}{c|}{RHF orbitals} & \multicolumn{3}{|c|}{NO orbitals} \\
\hline
CAS size - virtual orbs. & 6 & 12 & 16  & 6 & 12 & 16  \\ 
\hline
\multicolumn{7}{|l|}{\textbf{1.0 * Re, E(FCI) = -109.2804 Hartree}} \\
\hline
DMRG & 221.1 & 97.1 & 42.0 & 136.7 & 50.3 & 18.7   \\  
DMRG-DUCC(2) & 30.0 & 18.8 & 8.5 &  25.9 & 12.2 & 4.9  \\  
DMRG-DUCC(3) & 7.5 & 5.6 & 2.9   & 9.0 & 5.1 & 2.1   \\  
\hline
\multicolumn{7}{|l|}{\textbf{1.5 * Re, E(FCI) = -109.0680 Hartree}} \\
\hline
DMRG & 194.2 & 95.9 & 64.6 & 153.9 & 96.5 & 36.0  \\  
DMRG-DUCC(2) & 46.2 & 28.2 & 15.9 & 52.2 & 48.9 & 47.4  \\  
DMRG-DUCC(3) & 25.5 & 16.0 & 10.8 & 35.3 & 21.4 & 11.4  \\  
\hline
\multicolumn{7}{|l|}{\textbf{2.0 * Re, E(FCI) = -108.9715 Hartree}} \\
\hline
DMRG & 193.9 & 96.8 & 93.1 & 164.4 & 95.8 & 57.5  \\  
DMRG-DUCC(2) & 80.9 & 50.8 & 69.2 & 141.8 & 62.4 & 39.1  \\  
DMRG-DUCC(3) & 57.5 & 44.3 & 30.7 & 132.7 & 60.1 & 38.6  \\  

\hline
\end{tabular}
\caption{The energy differences $\Delta E = E(method)-E(FCI)$ for N$_2$ at different bond lengths. $\Delta E$ is in millihartrees (mH), relative to the DMRG energy in full space (E(FCI)) for each bond length. 
}
\label{tab1}
\end{table}

The results of the comparative study between DMRG with the bare restricted Hartree--Fock (RHF) reference and the DUCC-based variants for N$_2$ are presented in Tab. \ref{tab1}. We performed calculations for three different geometries corresponding to \(1.0 R_e\), \(1.5 R_e\), and \(2.0 R_e\) bond distances, where $R_e$ is the equilibrium distance $R_e = 2.068$ Bohrs. Also, we studied different active space sizes, specifically CAS(14,13)[6 active virtuals], CAS(14,19)[12 active virtuals], and CAS(14,23)[16 active virtuals], where no occupied orbitals were frozen and all occupied orbitals were taken into the CAS. We focused on errors relative to the FCI energies and reported the energy differences $\Delta E$ in millihartrees (mH) and defined as $\Delta E_{\rm method} = E_{\rm method}-E_{\rm FCI}$. We also compare two different set of orbitals, RHF and CCSD Natural orbitals (NO). 

At equilibrium, \(\text{N}_2\) is the single-reference system, where dynamical correlations dominate the correlation energy. The inclusion of the dynamical correlation via DUCC significantly improves the accuracy. For the smallest active space, DMRG with the bare RHF Hamiltonian gives an error of 221.1 mH relative to FCI, which is significantly reduced by including dynamical correlation with the DUCC Hamiltonians with the DMRG-DUCC(3) approach, reducing the error to only 7.5 mH. As the active space gets larger, these errors systematically reduce. As the bond is stretched, static correlation plays a significant role and at \(2.0 R_e\) is a dominant factor of the electronic structure. It is known that CCSD and CCSD(T) methods break down and give wrong (overestimated) energies with respect to FCI at the stretch geometries. Despite the underlying CCSD breaking down at these stretch bond lengths, DMRG with the DUCC effective Hamiltonians still captures a significant portion of the correlation outside of the active space and, in all cases, significantly improves energies compared to DMRG calculations with the bare Hamiltonian in the active space. 

Since CCSD is well-behaved at equilibrium for N$_2$, the CCSD natural orbitals generally help reduce the errors relative to FCI. In this case, natural orbitals from the CCSD one-particle density matrix already incorporate dynamic electron correlation effects into the smaller space around the single reference. The downfolding procedure balances out the shift in correlation from RHF orbitals to CCSD natural orbitals, and the change in energies is less significant than the changes for the calculations with the bare Hamiltonian in the active spaces.  However, as the bond is stretched and static correlation increases, critical configurations can be underrepresented when using natural orbitals of an insufficient method. 
In other words, transforming RHF orbitals to natural orbitals through a single-reference approach can concentrate the captured dynamical correlation within a limited set of (internal) orbitals, leaving the external space only partially connected to the internal space. Consequently, the DUCC downfolding technique cannot adequately capture the remaining external correlations. The CCSD method before downfolding in NO orbitals does not converge; however, DMRG-DUCC methods using CCSD unconverged amplitudes are still giving improved energy differences compared to the bare Hamiltonian in the same active space. 
The RHF orbitals provide an unbiased natural basis, and going forward, we recommend RHF orbitals for general use when downfolding correlation from external space. If one wants to improve the results with the DUCC effective Hamiltonians, they can switch to a natural orbital basis, provided the method for generating the natural orbitals can reliably describe the system.

To gain insight into how the DUCC downfolding affects the correlation structure, we investigated changes in natural orbital occupation numbers from DMRG calculations, single-orbital entropies, and mutual information. The largest changes are connected with the most correlated orbitals, numbered 5-10, because they are involved in the triple bond. 

\begin{figure}[htbp]
    \centering
    \includegraphics[width=1.0\textwidth]{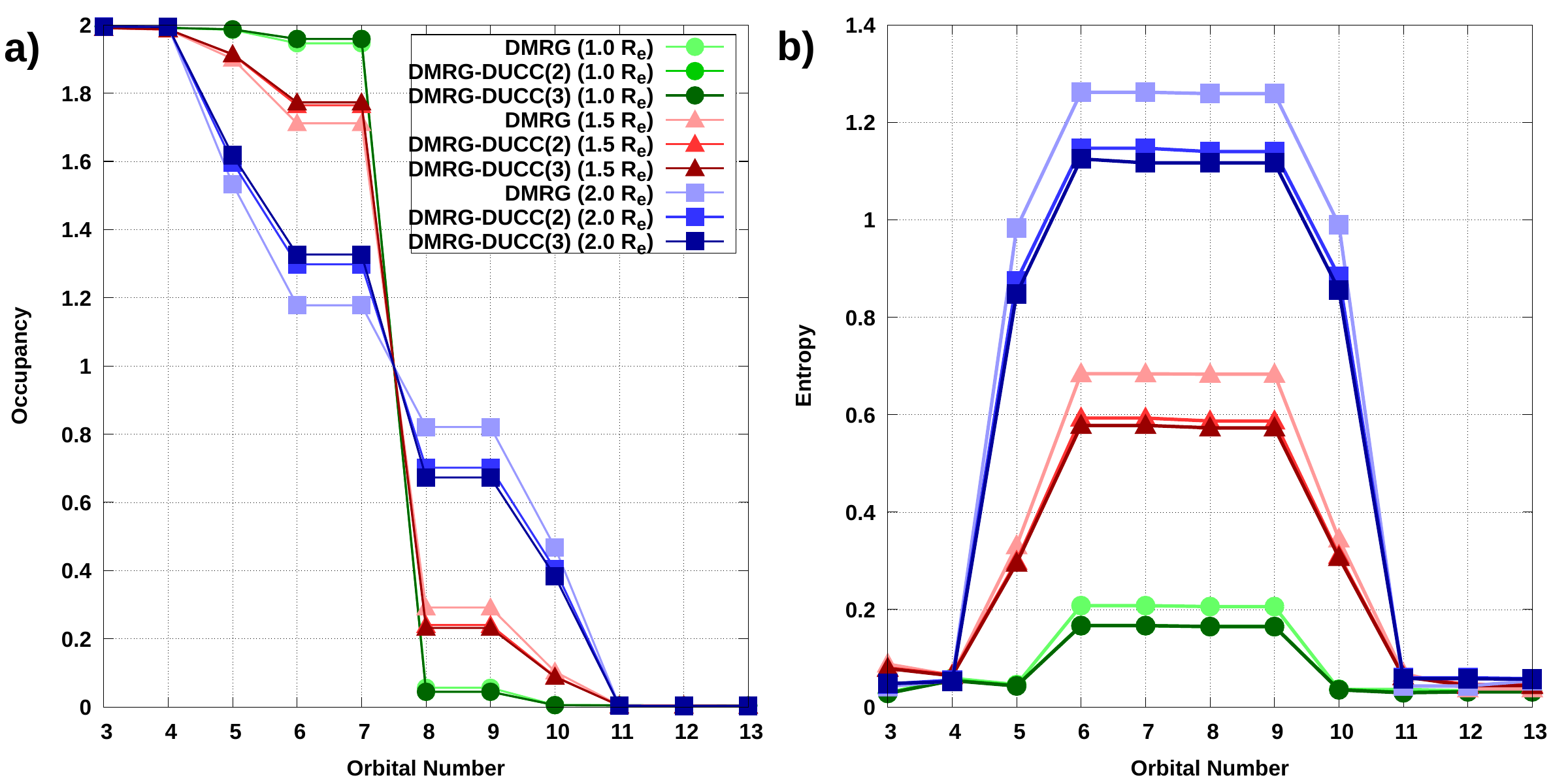}
    \caption{a) Natural occupation numbers and b) single-orbital entropies for the N$_2$ molecule, RHF orbitals and 13 orbital (6 virtual) active space. The two core orbitals are not shown.}
    \label{fig:occ}
\end{figure}

The natural orbital occupation numbers and single-orbital entropies are presented in Fig.~\ref{fig:occ}. The first two orbitals were excluded from the analysis because they correspond to core 1$s$ orbitals.  
At \( r = 1.0 R_e \), the differences between DMRG and the DUCC-based methods are very small, the occupation for two highest occupied and two lowest virtual orbitals (responsible for $\pi$-bonds) is approx. 1.95 and 0.05.
As the bond stretches to \( r = 1.5 R_e \) or \( r = 2.0 R_e \), the effects of static correlation become more significant, and the downfolding approaches consistently show an increase in the occupation of occupied orbitals and a reduction in virtual occupations compared to DMRG. This suggests that the DUCC downfolding leads to a stronger closed-shell character of the singlet state, consistent with previous studies, such as those on oligoacenes \cite{Lee2017, Schriber2018}, which observed that pure AS methods tend to overestimate radical character. However, incorporating dynamical correlation helps correct this by mitigating the dominance of static correlation effects.


The single-orbital entropies for RHF orbitals and CAS(14,13) are shown in Fig.\ref{fig:occ}a. At \( r = 1.0 R_e \), the entropies remain low across all methods, indicating a weak correlation in the system. However, as the bond is stretched to \( r = 1.5 R_e \) and \( r = 2.0 R_e \), entropies show a marked increase, highlighting the growing correlation effects in the system. However, the DUCC(2) and DUCC(3) downfolding methods exhibit slightly lower $s_i$ entropies than those of the bare Hamiltonian. This trend is in line with the observed effect in occupation numbers.

\begin{figure}[htbp]
    \centering
    \includegraphics[width=0.8\textwidth]{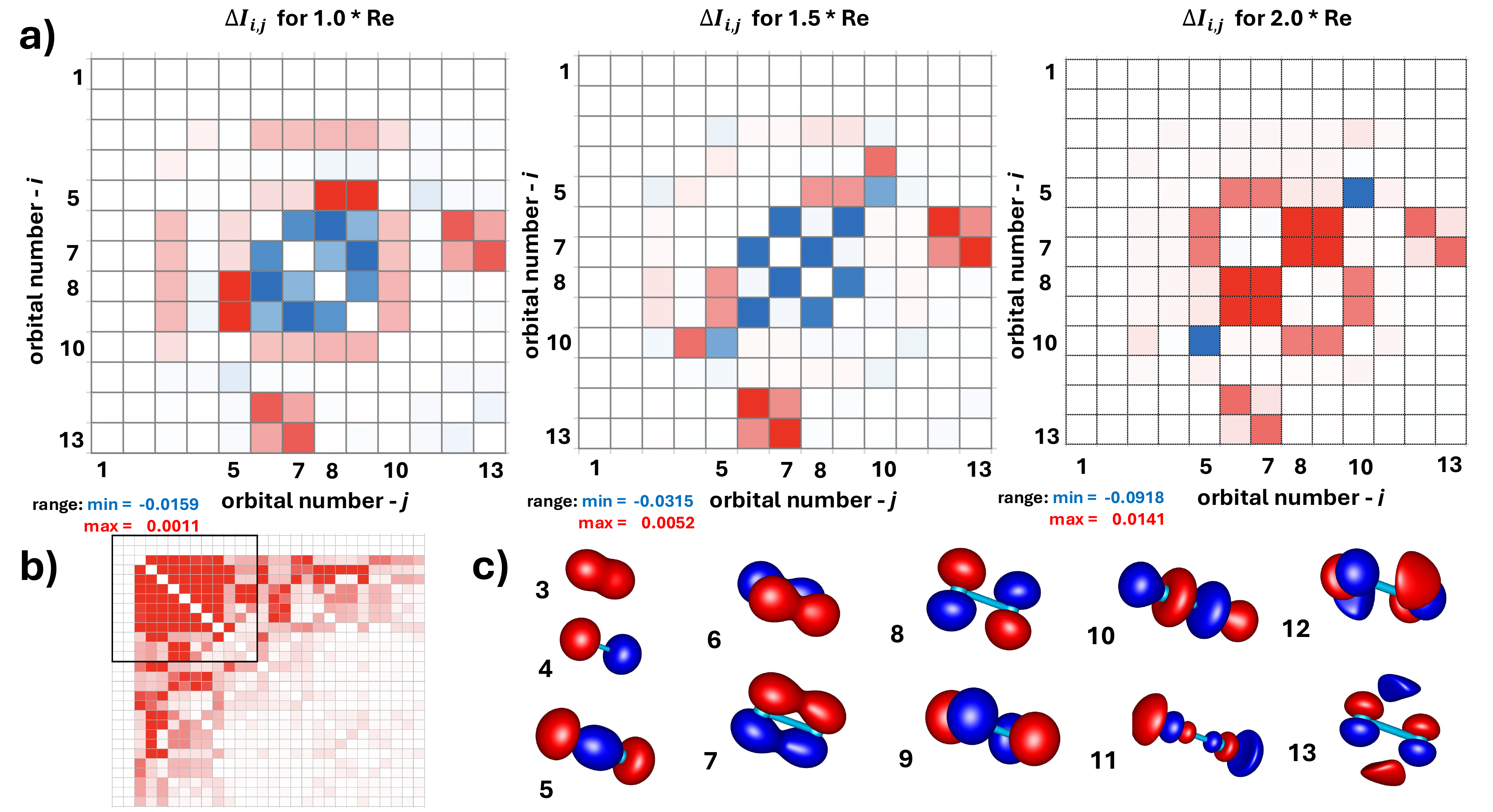}
    \caption{a) The mutual information difference $\Delta I_{i,j}$ heat map for the N$_2$ molecule. Results are for the RHF orbitals and the CAS(14,13) orbital space. The difference $\Delta I_{i,j}$ is computed as $I_{(\rm DMRG-DUCC(3))}-I_{(\rm DMRG)}$. The red color indicates an increase in $I_{i,j}$ for the DMRG-DUCC(3) method, and the blue color means a decrease in $I_{i,j}$. b) Heatmap  of $I_{i,j}$ obtained for full orbital space CAS(14,28) and \( r = 1.5 R_e \), solid red color is 90th percentile of $\Delta I_{i,j}$. The black frame corresponds to CAS(14,13) orbital space. c) Orbitals in CAS(14,13) space at \( r = 1.5 R_e \); orbitals numbered 1-7 are occupied, and 8-13 are virtual.}
    \label{fig:n2_mi}
\end{figure}

Fig. \ref{fig:n2_mi} shows the heat maps of the mutual information differences $\Delta I_{i,j} = I_{(\rm DMRG-DUCC(3))}-I_{(\rm DMRG)}$ in CAS(14,13) for the three bond distances. Each heat map represents the pairwise mutual information change for pair of orbitals $i,j$, with the color intensity reflecting the magnitude of the difference between the DMRG-DUCC(3) and DMRG method. Blue color indicates pairs for which $I_{i,j}$ is larger in DMRG, while red color highlights pairs for which $I_{i,j}$(DMRG-DUCC(3)) is larger.

At \( r = 1.0 R_e \), the $\Delta I_{i,j}$ heat map (left panel) exhibits an increase (red) in correlation captured by DMRG-DUCC(3) for orbitals 3, 5, 10, 12 and 13, and decrease (blue) for 
orbitals 6, 7, 8 and 9. $\Delta I_{i,j}$ is between -0.0159 and 0.0011. As discussed before, including the dynamical correlation in the AS enhances the single-reference character, which can be seen in negative $\Delta I_{i,j}$ for $\pi$-bond orbitals. On the other hand, we can see a stronger correlation for pairs participating in dynamical correlation contributions. 
For \( r = 1.5 R_e \) (middle panel), the differences between DMRG-DUCC(3) and DMRG become more significant. The heat map shows a decrease of $I_{i,j}$ in $\pi$-bond orbital region and an increase for lower and higher orbitals, consistently with the previous case. The most negative $\Delta I_{i,j}=-0.032$ is two-times larger than for \( r = 1.0 R_e \) case, the most positive value $\Delta I_{i,j}=0.005$ is five-times larger. The more substantial differences at this bond length are consistent with the expected increase in static correlation effects as the bond weakens. At \( r = 2.0 R_e \) case (right panel), $\Delta I_{i,j}$ shows significant variation, with dominating red regions. This indicates that DMRG-DUCC(3) captures substantially more correlation than DMRG alone, which is given by much stronger correlation in the external space. 




\begin{table}[ht]
\centering
\small 
\setlength{\tabcolsep}{6pt} 
\renewcommand{\arraystretch}{1.2} 
\begin{tabular}{|c|c|c|c|}
\hline
\textbf{} & \textbf{DMRG} & \textbf{DMRG-} & \textbf{DMRG-} \\

\textbf{} &  & \textbf{DUCC(2)}  & \textbf{DUCC(3)} \\
\hline
CAS(42,32)  & -230.8503 & -231.7796 & -231.8809  
\\
\hline
CAS(42,36)  & -230.8764 & -231.7809  & -231.8794 
\\
\hline
CAS(42,39)  & -230.8962 & -231.7822 & -231.8789  
\\
\hline
CAS(42,52)  & -231.0400 & -231.7912 & -231.8713 
\\
\hline
\end{tabular}
\caption{Total energies for the lowest singlet state of benzene obtained by DMRG and DUCC methods with various active space sizes (RHF orbital basis). CAS(X,Y) indicates the complete active space with X electrons in Y orbitals. Energies are reported in Hartree.}
\label{tab2}
\end{table}

The singlet ground state of benzene has a purely single-reference character with a significant energy gap between the HOMO/LUMO. However, the dynamical correlation effect is strong, so it could serve as a nontrivial testing system without the influence of static-like correlations. We selected four AS sizes, ranging from 32 to 52 orbitals. 
The energies shown in Tab.~\ref{tab2} demonstrate several trends. For all active space sizes, the utilization of DUCC(2) and DUCC(3) downfolding significantly improves upon the DMRG results with the bare RHF Hamiltonian, consistently yielding lower energies. The DUCC(3) approach introduces a notable improvement over DUCC(2). For reference, the all-electron CCSD and CCSD(T) energies are -231.84857 H and -231.90190 H, respectively. Because DMRG brings in higher-order correlations in the active space, the fact the results with the DUCC(3) Hamiltonians lie between these two values indicates that DMRG calculations with downfolded Hamiltonians are capable of providing energies comparable to full-orbital-space high-order electronic structure methods.  
For the DUCC(2) and DUCC(3) cases, the energies are relatively stable for all active spaces, varying less than 12 mH from the 32 to 52 orbital calculations, as compared to the nearly 190 mH range from DMRG with the bare RHF Hamiltonian in the same active spaces. 

\begin{figure}[htbp]
    \centering
    \includegraphics[width=0.6\textwidth]{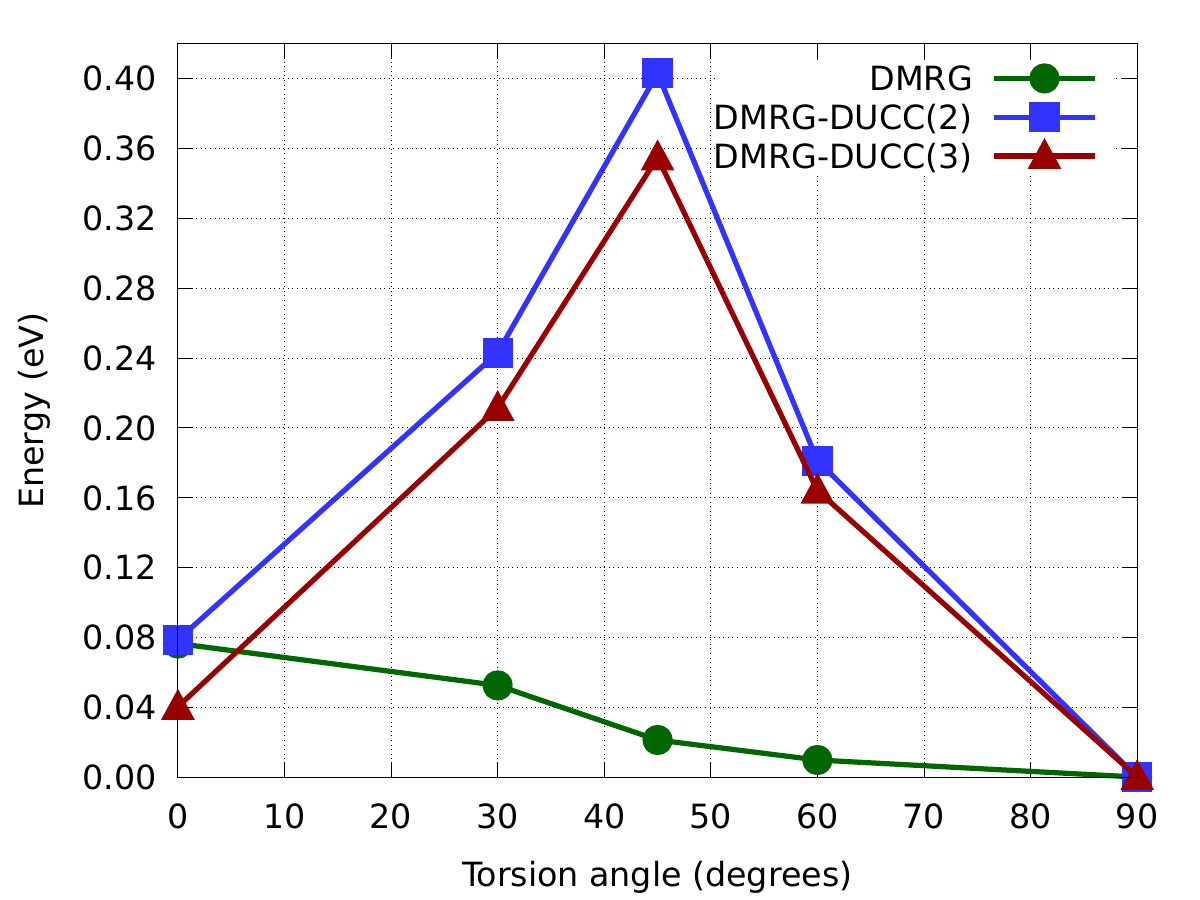}
    \caption{Relative energies in eV for TME (torsion angles 0, 30, 45 60, 90) obtained by DMRG and DUCC methods, with respect to the energy for $\alpha=90$.}
    \label{fig:6}
\end{figure}

The results for TME with CAS(44,34) are depicted in Fig.~\ref{fig:6}. In recent works\cite{Pozun2013, doi:10.1021/acs.jctc.8b00022} has been shown that the singlet PES has a maximum for torsion angle 45 degrees. However, a proper description 
requires an accurate account of the dynamical correlation effects; therefore, DMRG underestimates the correlation energy even in larger active spaces.
In our calculations, the DMRG PES maximum is $E_{\rm DMRG}(0)$, DMRG-DUCC methods correctly put $E(45)$ above $E(0)$ and $E(90)$, in the case of DMRG-DUCC(2) the twisting energy barrier is 0.35 eV and 0.4 eV for DMRG-DUCC(3). 

In summary, we have shown that the DUCC approach with the DMRG method can create a new, very robust, and accurate method for capturing both dynamical and static correlations using an acceptable active space size.
The DUCC approach projects the dynamical correlation from the external space to AS, which, in combination with DMRG, allows to achieve high accuracy even in cases where only dynamical correlation is dominant. 
We also studied the effect of DUCC downfolding on the correlation in AS. This is a crucial step for understanding the applicability and accuracy of the downfolding technique in future large-scale studies. Using single-orbital entropies, natural orbital occupancies, and changes in mutual information, we found that DUCC downfolding suppresses the multireference character and enhances the single-reference character in AS, while the correlation energy obtained from AS is substantially increased. 


\section{Acknowledgement}
This material is based upon work supported by the ``Transferring exascale computational chemistry to cloud computing environment and emerging hardware technologies (TEC$^4$)''  project, which is funded by the U.S. Department of Energy, Office of Science, Office of Basic Energy Sciences, the Division of Chemical Sciences, Geosciences, and Biosciences (under FWP 82037).  
 This work used resources from the Pacific Northwest National Laboratory (PNNL).
PNNL is operated by Battelle for the U.S. Department of Energy under Contract DE-AC05-76RL01830. LV and JB also acknowledge financial support from the Czech Science Foundation (grant no. 23-05486S), the Ministry of
Education, Youth and Sports of the Czech Republic through the e-INFRA CZ (ID:90254), and the Advanced Multiscale Materials for Key Enabling Technologies project, supported by the Ministry of Education, Youth, and Sports of the Czech Republic. Project No. CZ.02.01.01/00/22\_008/0004558, Co-funded by the European Union.

\bibliography{references}

\providecommand{\latin}[1]{#1}
\makeatletter
\providecommand{\doi}
  {\begingroup\let\do\@makeother\dospecials
  \catcode`\{=1 \catcode`\}=2 \doi@aux}
\providecommand{\doi@aux}[1]{\endgroup\texttt{#1}}
\makeatother
\providecommand*\mcitethebibliography{\thebibliography}
\csname @ifundefined\endcsname{endmcitethebibliography}  {\let\endmcitethebibliography\endthebibliography}{}
\begin{mcitethebibliography}{45}
\providecommand*\natexlab[1]{#1}
\providecommand*\mciteSetBstSublistMode[1]{}
\providecommand*\mciteSetBstMaxWidthForm[2]{}
\providecommand*\mciteBstWouldAddEndPuncttrue
  {\def\EndOfBibitem{\unskip.}}
\providecommand*\mciteBstWouldAddEndPunctfalse
  {\let\EndOfBibitem\relax}
\providecommand*\mciteSetBstMidEndSepPunct[3]{}
\providecommand*\mciteSetBstSublistLabelBeginEnd[3]{}
\providecommand*\EndOfBibitem{}
\mciteSetBstSublistMode{f}
\mciteSetBstMaxWidthForm{subitem}{(\alph{mcitesubitemcount})}
\mciteSetBstSublistLabelBeginEnd
  {\mcitemaxwidthsubitemform\space}
  {\relax}
  {\relax}

\bibitem[Coester(1958)]{coester58_421}
Coester,~F. Bound States of a Many-Particle System. \emph{Nucl. Phys.} \textbf{1958}, \emph{7}, 421--424\relax
\mciteBstWouldAddEndPuncttrue
\mciteSetBstMidEndSepPunct{\mcitedefaultmidpunct}
{\mcitedefaultendpunct}{\mcitedefaultseppunct}\relax
\EndOfBibitem
\bibitem[Coester and Kummel(1960)Coester, and Kummel]{coester60_477}
Coester,~F.; Kummel,~H. Short-Range Correlations in Nuclear Wave Functions. \emph{Nucl. Phys.} \textbf{1960}, \emph{17}, 477--485\relax
\mciteBstWouldAddEndPuncttrue
\mciteSetBstMidEndSepPunct{\mcitedefaultmidpunct}
{\mcitedefaultendpunct}{\mcitedefaultseppunct}\relax
\EndOfBibitem
\bibitem[{\v C}{\'\i}{\v z}ek(1966)]{cizek66_4256}
{\v C}{\'\i}{\v z}ek,~J. On the Correlation Problem in Atomic and Molecular Systems. Calculation of Wavefunction Components in Ursell-Type Expansion Using Quantum-Field Theoretical Methods. \emph{J. Chem. Phys.} \textbf{1966}, \emph{45}, 4256--4266\relax
\mciteBstWouldAddEndPuncttrue
\mciteSetBstMidEndSepPunct{\mcitedefaultmidpunct}
{\mcitedefaultendpunct}{\mcitedefaultseppunct}\relax
\EndOfBibitem
\bibitem[Paldus \latin{et~al.}(1972)Paldus, {\v{C}}{\'\i}{\v{z}}ek, and Shavitt]{paldus1972correlation}
Paldus,~J.; {\v{C}}{\'\i}{\v{z}}ek,~J.; Shavitt,~I. Correlation Problems in Atomic and Molecular Systems. IV. Extended Coupled-Pair Many-Electron Theory and Its Application to the BH$_3$ Molecule. \emph{Phys. Rev. A} \textbf{1972}, \emph{5}, 50\relax
\mciteBstWouldAddEndPuncttrue
\mciteSetBstMidEndSepPunct{\mcitedefaultmidpunct}
{\mcitedefaultendpunct}{\mcitedefaultseppunct}\relax
\EndOfBibitem
\bibitem[Purvis and Bartlett(1982)Purvis, and Bartlett]{purvis82_1910}
Purvis,~G.~D.; Bartlett,~R.~J. A Full Coupled-Cluster Singles and Doubles Model: The Inclusion of Disconnected Triples. \emph{J. Chem. Phys.} \textbf{1982}, \emph{76}, 1910--1918\relax
\mciteBstWouldAddEndPuncttrue
\mciteSetBstMidEndSepPunct{\mcitedefaultmidpunct}
{\mcitedefaultendpunct}{\mcitedefaultseppunct}\relax
\EndOfBibitem
\bibitem[Paldus and Li(1999)Paldus, and Li]{paldus07}
Paldus,~J.; Li,~X. A Critical Assessment of Coupled Cluster Method in Quantum Chemistry. \emph{Adv. Chem. Phys.} \textbf{1999}, \emph{110}, 1--175\relax
\mciteBstWouldAddEndPuncttrue
\mciteSetBstMidEndSepPunct{\mcitedefaultmidpunct}
{\mcitedefaultendpunct}{\mcitedefaultseppunct}\relax
\EndOfBibitem
\bibitem[Crawford and Schaefer(2000)Crawford, and Schaefer]{crawford2000introduction}
Crawford,~T.~D.; Schaefer,~H.~F. An introduction to coupled cluster theory for computational chemists. \emph{Reviews in computational chemistry} \textbf{2000}, \emph{14}, 33--136\relax
\mciteBstWouldAddEndPuncttrue
\mciteSetBstMidEndSepPunct{\mcitedefaultmidpunct}
{\mcitedefaultendpunct}{\mcitedefaultseppunct}\relax
\EndOfBibitem
\bibitem[Bartlett and Musia\l(2007)Bartlett, and Musia\l]{bartlett_rmp}
Bartlett,~R.~J.; Musia\l,~M. Coupled-Cluster Theory in Quantum Chemistry. \emph{Rev. Mod. Phys.} \textbf{2007}, \emph{79}, 291--352\relax
\mciteBstWouldAddEndPuncttrue
\mciteSetBstMidEndSepPunct{\mcitedefaultmidpunct}
{\mcitedefaultendpunct}{\mcitedefaultseppunct}\relax
\EndOfBibitem
\bibitem[Neuscamman \latin{et~al.}(2010)Neuscamman, Yanai, and Chan]{neuscamman_2010_irpc}
Neuscamman,~E.; Yanai,~T.; Chan,~G. K.-L. A review of canonical transformation theory. \emph{Int. Rev. Phys. Chem.} \textbf{2010}, \emph{29}, 231--271\relax
\mciteBstWouldAddEndPuncttrue
\mciteSetBstMidEndSepPunct{\mcitedefaultmidpunct}
{\mcitedefaultendpunct}{\mcitedefaultseppunct}\relax
\EndOfBibitem
\bibitem[Roemelt \latin{et~al.}(2016)Roemelt, Guo, and Chan]{Roemelt2016}
Roemelt,~M.; Guo,~S.; Chan,~G. K.-L. A projected approximation to strongly contracted N-electron valence perturbation theory for {DMRG} wavefunctions. \emph{J. Chem. Phys.} \textbf{2016}, \emph{144}, 204113\relax
\mciteBstWouldAddEndPuncttrue
\mciteSetBstMidEndSepPunct{\mcitedefaultmidpunct}
{\mcitedefaultendpunct}{\mcitedefaultseppunct}\relax
\EndOfBibitem
\bibitem[Freitag \latin{et~al.}(2017)Freitag, Knecht, Angeli, and Reiher]{Freitag2017}
Freitag,~L.; Knecht,~S.; Angeli,~C.; Reiher,~M. Multireference Perturbation Theory with Cholesky Decomposition for the Density Matrix Renormalization Group. \emph{J. Chem. Theory Comput.} \textbf{2017}, \emph{13}, 451--459\relax
\mciteBstWouldAddEndPuncttrue
\mciteSetBstMidEndSepPunct{\mcitedefaultmidpunct}
{\mcitedefaultendpunct}{\mcitedefaultseppunct}\relax
\EndOfBibitem
\bibitem[Sharma and Chan(2014)Sharma, and Chan]{sharma_2014c}
Sharma,~S.; Chan,~G. A flexible multi-reference perturbation theory by minimizing the Hylleraas functional with matrix product states. \emph{J. Chem. Phys.} \textbf{2014}, \emph{141}, 111101\relax
\mciteBstWouldAddEndPuncttrue
\mciteSetBstMidEndSepPunct{\mcitedefaultmidpunct}
{\mcitedefaultendpunct}{\mcitedefaultseppunct}\relax
\EndOfBibitem
\bibitem[Veis \latin{et~al.}(2016)Veis, Antal{\'{\i}}k, Brabec, Neese, \"{O}rs Legeza, and Pittner]{Veis2016}
Veis,~L.; Antal{\'{\i}}k,~A.; Brabec,~J.; Neese,~F.; \"{O}rs Legeza; Pittner,~J. Coupled Cluster Method with Single and Double Excitations Tailored by Matrix Product State Wave Functions. \emph{J. Phys. Chem. Lett.} \textbf{2016}, \emph{7}, 4072--4078\relax
\mciteBstWouldAddEndPuncttrue
\mciteSetBstMidEndSepPunct{\mcitedefaultmidpunct}
{\mcitedefaultendpunct}{\mcitedefaultseppunct}\relax
\EndOfBibitem
\bibitem[Beran \latin{et~al.}(2021)Beran, Matoušek, Hapka, Pernal, and Veis]{Beran2021}
Beran,~P.; Matoušek,~M.; Hapka,~M.; Pernal,~K.; Veis,~L. Density Matrix Renormalization Group with Dynamical Correlation via Adiabatic Connection. \emph{Journal of Chemical Theory and Computation} \textbf{2021}, \emph{17}, 7575–7585\relax
\mciteBstWouldAddEndPuncttrue
\mciteSetBstMidEndSepPunct{\mcitedefaultmidpunct}
{\mcitedefaultendpunct}{\mcitedefaultseppunct}\relax
\EndOfBibitem
\bibitem[Kowalski(2021)]{kowalski2021dimensionality}
Kowalski,~K. Dimensionality reduction of the many-body problem using coupled-cluster subsystem flow equations: Classical and quantum computing perspective. \emph{Phys. Rev. A} \textbf{2021}, \emph{104}, 032804\relax
\mciteBstWouldAddEndPuncttrue
\mciteSetBstMidEndSepPunct{\mcitedefaultmidpunct}
{\mcitedefaultendpunct}{\mcitedefaultseppunct}\relax
\EndOfBibitem
\bibitem[Bauman and Kowalski(2022)Bauman, and Kowalski]{bauman2022coupled}
Bauman,~N.~P.; Kowalski,~K. Coupled Cluster Downfolding Theory: towards universal many-body algorithms for dimensionality reduction of composite quantum systems in chemistry and materials science. \emph{Mater. Theory} \textbf{2022}, \emph{6}, 1--19\relax
\mciteBstWouldAddEndPuncttrue
\mciteSetBstMidEndSepPunct{\mcitedefaultmidpunct}
{\mcitedefaultendpunct}{\mcitedefaultseppunct}\relax
\EndOfBibitem
\bibitem[Kowalski and Bauman(2023)Kowalski, and Bauman]{kowalski2023quantum}
Kowalski,~K.; Bauman,~N.~P. Quantum flow algorithms for simulating many-body systems on quantum computers. \emph{Physical Review Letters} \textbf{2023}, \emph{131}, 200601\relax
\mciteBstWouldAddEndPuncttrue
\mciteSetBstMidEndSepPunct{\mcitedefaultmidpunct}
{\mcitedefaultendpunct}{\mcitedefaultseppunct}\relax
\EndOfBibitem
\bibitem[Schollw\"ock(2005)]{schollwock_2005}
Schollw\"ock,~U. \emph{Rev. Mod. Phys.} \textbf{2005}, \emph{77}, 259--315\relax
\mciteBstWouldAddEndPuncttrue
\mciteSetBstMidEndSepPunct{\mcitedefaultmidpunct}
{\mcitedefaultendpunct}{\mcitedefaultseppunct}\relax
\EndOfBibitem
\bibitem[White and Martin(1999)White, and Martin]{White1999}
White,~S.~R.; Martin,~R.~L. Ab Initio Quantum Chemistry using the Density Matrix Renormalization Group. \emph{J. Chem. Phys.} \textbf{1999}, \emph{110}, 4127--4130\relax
\mciteBstWouldAddEndPuncttrue
\mciteSetBstMidEndSepPunct{\mcitedefaultmidpunct}
{\mcitedefaultendpunct}{\mcitedefaultseppunct}\relax
\EndOfBibitem
\bibitem[Chan and Sharma(2011)Chan, and Sharma]{chan_review}
Chan,~G. K.-L.; Sharma,~S. The Density Matrix Renormalization Group in Quantum Chemistry. \emph{Ann. Rev. Phys. Chem.} \textbf{2011}, \emph{62}, 465--481\relax
\mciteBstWouldAddEndPuncttrue
\mciteSetBstMidEndSepPunct{\mcitedefaultmidpunct}
{\mcitedefaultendpunct}{\mcitedefaultseppunct}\relax
\EndOfBibitem
\bibitem[Wouters and Van~Neck(2014)Wouters, and Van~Neck]{wouters_review}
Wouters,~S.; Van~Neck,~D. The Density Matrix Renormalization Group for Ab Initio Quantum Chemistry. \emph{Eur. Phys. J. D} \textbf{2014}, \emph{68}\relax
\mciteBstWouldAddEndPuncttrue
\mciteSetBstMidEndSepPunct{\mcitedefaultmidpunct}
{\mcitedefaultendpunct}{\mcitedefaultseppunct}\relax
\EndOfBibitem
\bibitem[Szalay \latin{et~al.}(2015)Szalay, Pfeffer, Murg, Barcza, Verstraete, Schneider, and \"{O}rs Legeza]{Szalay2015}
Szalay,~S.; Pfeffer,~M.; Murg,~V.; Barcza,~G.; Verstraete,~F.; Schneider,~R.; \"{O}rs Legeza Tensor Product Methods and Entanglement Optimization for Ab Initio Quantum Chemistry. \emph{Int. J. Quant. Chem.} \textbf{2015}, \emph{115}, 1342--1391\relax
\mciteBstWouldAddEndPuncttrue
\mciteSetBstMidEndSepPunct{\mcitedefaultmidpunct}
{\mcitedefaultendpunct}{\mcitedefaultseppunct}\relax
\EndOfBibitem
\bibitem[Baiardi and Reiher(2020)Baiardi, and Reiher]{reiher_perspective}
Baiardi,~A.; Reiher,~M. The Density Matrix Renormalization Group in Chemistry and Molecular Physics: Recent Developments and new Challenges. \emph{J. Chem. Phys.} \textbf{2020}, \emph{152}, 040903\relax
\mciteBstWouldAddEndPuncttrue
\mciteSetBstMidEndSepPunct{\mcitedefaultmidpunct}
{\mcitedefaultendpunct}{\mcitedefaultseppunct}\relax
\EndOfBibitem
\bibitem[White(1992)]{White1992}
White,~S.~R. Density matrix formulation for quantum renormalization groups. \emph{Phys. Rev. Lett.} \textbf{1992}, \emph{69}, 2863\relax
\mciteBstWouldAddEndPuncttrue
\mciteSetBstMidEndSepPunct{\mcitedefaultmidpunct}
{\mcitedefaultendpunct}{\mcitedefaultseppunct}\relax
\EndOfBibitem
\bibitem[White(1993)]{White-1993}
White,~S.~R. Density-matrix algorithms for quantum renormalization groups. \emph{Phys. Rev. B} \textbf{1993}, \emph{48}, 10345--10356\relax
\mciteBstWouldAddEndPuncttrue
\mciteSetBstMidEndSepPunct{\mcitedefaultmidpunct}
{\mcitedefaultendpunct}{\mcitedefaultseppunct}\relax
\EndOfBibitem
\bibitem[Schollw{\"o}ck(2011)]{Schollwock2011}
Schollw{\"o}ck,~U. The density-matrix renormalization group in the age of matrix product states. \emph{Ann. Phys.} \textbf{2011}, \emph{326}, 96--192\relax
\mciteBstWouldAddEndPuncttrue
\mciteSetBstMidEndSepPunct{\mcitedefaultmidpunct}
{\mcitedefaultendpunct}{\mcitedefaultseppunct}\relax
\EndOfBibitem
\bibitem[Keller \latin{et~al.}(2015)Keller, Dolfi, Troyer, and Reiher]{keller_2015}
Keller,~S.; Dolfi,~M.; Troyer,~M.; Reiher,~M. \emph{J. Chem. Phys.} \textbf{2015}, \emph{143}, 244118\relax
\mciteBstWouldAddEndPuncttrue
\mciteSetBstMidEndSepPunct{\mcitedefaultmidpunct}
{\mcitedefaultendpunct}{\mcitedefaultseppunct}\relax
\EndOfBibitem
\bibitem[Keller and Reiher(2016)Keller, and Reiher]{keller_2016}
Keller,~S.; Reiher,~M. Spin-adapted Matrix Product States and Operators. \emph{J. Chem. Phys.} \textbf{2016}, \emph{144}, 134101\relax
\mciteBstWouldAddEndPuncttrue
\mciteSetBstMidEndSepPunct{\mcitedefaultmidpunct}
{\mcitedefaultendpunct}{\mcitedefaultseppunct}\relax
\EndOfBibitem
\bibitem[Chan \latin{et~al.}(2016)Chan, Keselman, Nakatani, Li, and White]{chan_mpo}
Chan,~G. K.-L.; Keselman,~A.; Nakatani,~N.; Li,~Z.; White,~S.~R. \emph{J. Chem. Phys.} \textbf{2016}, \emph{145}, 014102\relax
\mciteBstWouldAddEndPuncttrue
\mciteSetBstMidEndSepPunct{\mcitedefaultmidpunct}
{\mcitedefaultendpunct}{\mcitedefaultseppunct}\relax
\EndOfBibitem
\bibitem[Legeza and S\'olyom(2003)Legeza, and S\'olyom]{legeza_2003b}
Legeza,~{\"O}.; S\'olyom,~J. Optimizing the density-matrix renormalization group method using quantum information entropy. \emph{Phys. Rev. B} \textbf{2003}, \emph{68}, 195116\relax
\mciteBstWouldAddEndPuncttrue
\mciteSetBstMidEndSepPunct{\mcitedefaultmidpunct}
{\mcitedefaultendpunct}{\mcitedefaultseppunct}\relax
\EndOfBibitem
\bibitem[Rissler \latin{et~al.}(2006)Rissler, Noack, and White]{rissler_2006}
Rissler,~J.; Noack,~R.~M.; White,~S.~R. Measuring orbital interaction using quantum information theory. \emph{Chem. Phys.} \textbf{2006}, \emph{323}, 519 -- 531\relax
\mciteBstWouldAddEndPuncttrue
\mciteSetBstMidEndSepPunct{\mcitedefaultmidpunct}
{\mcitedefaultendpunct}{\mcitedefaultseppunct}\relax
\EndOfBibitem
\bibitem[Barcza \latin{et~al.}(2011)Barcza, Legeza, Marti, and Reiher]{barcza_2011}
Barcza,~G.; Legeza,~{\"O}.; Marti,~K.~H.; Reiher,~M. Quantum-information analysis of electronic states of different molecular structures. \emph{Phys. Rev. A} \textbf{2011}, \emph{83}, 012508\relax
\mciteBstWouldAddEndPuncttrue
\mciteSetBstMidEndSepPunct{\mcitedefaultmidpunct}
{\mcitedefaultendpunct}{\mcitedefaultseppunct}\relax
\EndOfBibitem
\bibitem[Barcza \latin{et~al.}(2015)Barcza, Noack, S\'{o}lyom, and Legeza]{barcza_2015}
Barcza,~G.; Noack,~R.~M.; S\'{o}lyom,~J.; Legeza,~{\"O}. \emph{Phys. Rev. B} \textbf{2015}, \emph{92}, 125140\relax
\mciteBstWouldAddEndPuncttrue
\mciteSetBstMidEndSepPunct{\mcitedefaultmidpunct}
{\mcitedefaultendpunct}{\mcitedefaultseppunct}\relax
\EndOfBibitem
\bibitem[Kowalski(2018)]{kowalski2018properties}
Kowalski,~K. Properties of coupled-cluster equations originating in excitation sub-algebras. \emph{J. Chem. Phys.} \textbf{2018}, \emph{148}, 094104\relax
\mciteBstWouldAddEndPuncttrue
\mciteSetBstMidEndSepPunct{\mcitedefaultmidpunct}
{\mcitedefaultendpunct}{\mcitedefaultseppunct}\relax
\EndOfBibitem
\bibitem[Kowalski(2023)]{kowalski2023sub}
Kowalski,~K. Sub-system self-consistency in coupled cluster theory. \emph{J. Chem. Phys.} \textbf{2023}, \emph{158}, 054101\relax
\mciteBstWouldAddEndPuncttrue
\mciteSetBstMidEndSepPunct{\mcitedefaultmidpunct}
{\mcitedefaultendpunct}{\mcitedefaultseppunct}\relax
\EndOfBibitem
\bibitem[Bauman \latin{et~al.}(2019)Bauman, Bylaska, Krishnamoorthy, Low, Wiebe, Granade, Roetteler, Troyer, and Kowalski]{bauman2019downfolding}
Bauman,~N.~P.; Bylaska,~E.~J.; Krishnamoorthy,~S.; Low,~G.~H.; Wiebe,~N.; Granade,~C.~E.; Roetteler,~M.; Troyer,~M.; Kowalski,~K. Downfolding of many-body Hamiltonians using active-space models: Extension of the sub-system embedding sub-algebras approach to unitary coupled cluster formalisms. \emph{J. Chem. Phys.} \textbf{2019}, \emph{151}, 014107\relax
\mciteBstWouldAddEndPuncttrue
\mciteSetBstMidEndSepPunct{\mcitedefaultmidpunct}
{\mcitedefaultendpunct}{\mcitedefaultseppunct}\relax
\EndOfBibitem
\bibitem[Piecuch \latin{et~al.}(1993)Piecuch, Oliphant, and Adamowicz]{pnl93}
Piecuch,~P.; Oliphant,~N.; Adamowicz,~L. A state-selective multireference coupled-cluster theory employing the single-reference formalism. \emph{J. Chem. Phys.} \textbf{1993}, \emph{99}, 1875--1900\relax
\mciteBstWouldAddEndPuncttrue
\mciteSetBstMidEndSepPunct{\mcitedefaultmidpunct}
{\mcitedefaultendpunct}{\mcitedefaultseppunct}\relax
\EndOfBibitem
\bibitem[Bauman and Kowalski(2022)Bauman, and Kowalski]{doublec2022}
Bauman,~N.~P.; Kowalski,~K. Coupled cluster downfolding methods: The effect of double commutator terms on the accuracy of ground-state energies. \emph{The Journal of Chemical Physics} \textbf{2022}, \emph{156}, 094106\relax
\mciteBstWouldAddEndPuncttrue
\mciteSetBstMidEndSepPunct{\mcitedefaultmidpunct}
{\mcitedefaultendpunct}{\mcitedefaultseppunct}\relax
\EndOfBibitem
\bibitem[Mutlu \latin{et~al.}(2023)Mutlu, Panyala, Gawande, Bagusetty, Glabe, Kim, Kowalski, Bauman, Peng, Pathak, \latin{et~al.} others]{mutlu2023tamm}
Mutlu,~E.; Panyala,~A.; Gawande,~N.; Bagusetty,~A.; Glabe,~J.; Kim,~J.; Kowalski,~K.; Bauman,~N.~P.; Peng,~B.; Pathak,~H.; others TAMM: Tensor algebra for many-body methods. \emph{The Journal of Chemical Physics} \textbf{2023}, \emph{159}\relax
\mciteBstWouldAddEndPuncttrue
\mciteSetBstMidEndSepPunct{\mcitedefaultmidpunct}
{\mcitedefaultendpunct}{\mcitedefaultseppunct}\relax
\EndOfBibitem
\bibitem[Brabec \latin{et~al.}(2021)Brabec, Brandejs, Kowalski, Xantheas, Legeza, and Veis]{molmps}
Brabec,~J.; Brandejs,~J.; Kowalski,~K.; Xantheas,~S.; Legeza,~{\"O}.; Veis,~L. Massively parallel quantum chemical density matrix renormalization group method. \emph{J. Comput. Chem.} \textbf{2021}, \emph{42}, 534--544\relax
\mciteBstWouldAddEndPuncttrue
\mciteSetBstMidEndSepPunct{\mcitedefaultmidpunct}
{\mcitedefaultendpunct}{\mcitedefaultseppunct}\relax
\EndOfBibitem
\bibitem[Lee \latin{et~al.}(2017)Lee, Small, Epifanovsky, and Head-Gordon]{Lee2017}
Lee,~J.; Small,~D.~W.; Epifanovsky,~E.; Head-Gordon,~M. Coupled-Cluster Valence-Bond Singles and Doubles for Strongly Correlated Systems: Block-Tensor Based Implementation and Application to Oligoacenes. \emph{Journal of Chemical Theory and Computation} \textbf{2017}, \emph{13}, 602–615\relax
\mciteBstWouldAddEndPuncttrue
\mciteSetBstMidEndSepPunct{\mcitedefaultmidpunct}
{\mcitedefaultendpunct}{\mcitedefaultseppunct}\relax
\EndOfBibitem
\bibitem[Schriber \latin{et~al.}(2018)Schriber, Hannon, Li, and Evangelista]{Schriber2018}
Schriber,~J.~B.; Hannon,~K.~P.; Li,~C.; Evangelista,~F.~A. A Combined Selected Configuration Interaction and Many-Body Treatment of Static and Dynamical Correlation in Oligoacenes. \emph{Journal of Chemical Theory and Computation} \textbf{2018}, \emph{14}, 6295–6305\relax
\mciteBstWouldAddEndPuncttrue
\mciteSetBstMidEndSepPunct{\mcitedefaultmidpunct}
{\mcitedefaultendpunct}{\mcitedefaultseppunct}\relax
\EndOfBibitem
\bibitem[Pozun \latin{et~al.}(2013)Pozun, Su, and Jordan]{Pozun2013}
Pozun,~Z.~D.; Su,~X.; Jordan,~K.~D. A Study on the Properties and Dynamics of the Title Compound. \emph{Journal of the American Chemical Society} \textbf{2013}, \emph{135}, 13862–13869\relax
\mciteBstWouldAddEndPuncttrue
\mciteSetBstMidEndSepPunct{\mcitedefaultmidpunct}
{\mcitedefaultendpunct}{\mcitedefaultseppunct}\relax
\EndOfBibitem
\bibitem[Veis \latin{et~al.}(2018)Veis, Antalík, Legeza, Alavi, and Pittner]{doi:10.1021/acs.jctc.8b00022}
Veis,~L.; Antalík,~A.; Legeza,~{\"O}.; Alavi,~A.; Pittner,~J. The Intricate Case of Tetramethyleneethane: A Full Configuration Interaction Quantum Monte Carlo Benchmark and Multireference Coupled Cluster Studies. \emph{Journal of Chemical Theory and Computation} \textbf{2018}, \emph{14}, 2439--2445, PMID: 29570291\relax
\mciteBstWouldAddEndPuncttrue
\mciteSetBstMidEndSepPunct{\mcitedefaultmidpunct}
{\mcitedefaultendpunct}{\mcitedefaultseppunct}\relax
\EndOfBibitem
\end{mcitethebibliography}

\end{document}